\documentclass[superscriptaddress, nofootinbib, twocolumn, amsmath,amssymb, aps, pra, notitlepage, longbibliography]{revtex4-1}

\usepackage{dcolumn}
\usepackage{bm}
\usepackage{epsfig}
\usepackage{graphicx}
\usepackage{latexsym}
\usepackage{amsfonts}
\usepackage{setspace}
\usepackage{graphicx}
\usepackage{amsmath}
\usepackage{verbatim}
\usepackage{color}
\usepackage{siunitx}
\usepackage{hyperref}
\usepackage{nccmath}
\usepackage{upgreek}
\usepackage{hyperref}

\begin{document}

\title{On-chip correlation-based Brillouin sensing: design, experiment and simulation}

\author{Atiyeh Zarifi$^{1,2,\ast}$, Birgit Stiller$^{1,2}$, Moritz Merklein$^{1,2}$, Yang Liu$^{1,2}$, Blair Morrison$^{1,2}$, Alvaro Casas-Bedoya$^{1,2}$, Ganghui Ren$^{3}$, Thach G. Nguyen$^{3}$, Khu Vu$^{4}$, Duk-Yong Choi$^{4}$, Arnan Mitchell$^{3}$, Stephen J. Madden$^{4}$ and Benjamin J. Eggleton$^{1,2}$\\
\small{ \textcolor{white}{blanc\\}
$^{1}$Institute of Photonics and Optical Science (IPOS), School of Physics, The University of Sydney, NSW 2006, Australia.\\
$^{2}$Sydney Nano Institute (Sydney Nano), The University of Sydney, NSW 2006, Australia.\\
$^{3}$School of Engineering, RMIT University, Melbourne, VIC 3001, Australia.\\
$^{4}$Laser Physics Centre, Research School of Physics and Engineering, Australian National University, Canberra, ACT 2601, Australia.\\
$^{\ast}$atiyeh.zarifi@sydney.edu.au}}

%\email{birgit.stiller@sydney.edu.au}
%\date{\today}

\begin{abstract}
Wavelength-scale SBS waveguides are enabling novel on-chip functionalities. The micro- and nano-scale SBS structures and the complexity of the SBS waveguides require a characterization technique to monitor the local geometry-dependent SBS responses along the waveguide. In this work, we experimentally demonstrate detection of longitudinal features down to \SI{200}{\um} on a silicon-chalcogenide waveguide using the Brillouin optical correlation domain analysis (BOCDA) technique. We provide extensive simulation and analysis of how multiple acoustic and optical modes and geometrical variations influence the Brillouin spectrum.

\end{abstract}

\maketitle

%\ocis{() ;
%      () ;
%      () .}

%\bibliography{bibliography,extra}
%\bibliographystyle{osajnl}

%%%%%%%%%%%%%%%%%%%%%%%%%%%%%%%%%%%%%%%%%%%%%%%%%%%%%%%%%%%%%%%%%%%%%%%%%%%%%%%%
%%%%%%%%%%%%%%%%%%%%%%%%%%%%%%%%%%%%%%%%%%%%%%%%%%%%%%%%%%%%%%%%%%%%%%%%%%%%%%%%

Stimulated Brillouin scattering (SBS) is an inelastic scattering process in which a pump photon interacts with an acoustic phonon and generates a Stokes photon. The generated photon is down-shifted from the pump frequency by the acoustic resonance frequency and its linewidth is dependent upon the acoustic phonon lifetime in the medium \cite{Agrawal}. 
The hypersonic (\SI{}{\GHz}) frequency shift resulting from the SBS process provides a bridge between electronics and photonics enabling powerful applications such as pure microwave sources \cite{Li2013,Merklein2016} and tunable radio frequency (RF) filters \cite{Marpaung2015a,Sancho2012a}. The narrow linewidth of the SBS process makes it suitable for Brillouin-based lasers \cite{Kabakova2013,Lee2012,Loh2015a,Otterstrom2018,Morrison2017} and frequency comb generation \cite{Buttner2014,Braje2009}. Furthermore, the difference between the light and sound velocity enables light storage applications in photonics waveguides 
\cite{Harrison2013,Dong2015a,Merklein2016b}. Finally, since the SBS frequency shift is an intrinsic characteristic of the medium, SBS has become an ideal sensing mechanism in optical fiber networks \cite{Hotate2013,Bao1993,Nikles1996b,Kurashima1990,Thevenaz1999,Cohen2014a,Zhang2018e,Song2006,Motil2016}.
Since the SBS response is sensitive to environmental variables such as temperature and strain, it has been adopted as a distributed sensing mechanism in long optical fibers to monitor critical structures such as buildings and bridges \cite{Thevenaz1999,Bao1993,Song2006}. 

The spatial resolution required in structural health monitoring ranges from a few meters to a few \SI{}{\cm} depending on the application \cite{Thevenaz1999,Thevenaz2010}, which can be achieved using a distributed SBS measurement such as Brillouin optical time domain analysis (BOTDA). This approach employs optical pump pulses whose duration determines the spatial resolution of the SBS response \cite{Kurashima1990}. However, for a pulse duration shorter than the acoustic lifetime, the SBS spectrum broadens and the gain reduces significantly \cite{Bao1993}. Therefore, this approach is limited by the phonon lifetime in optical fibers - approximately \SI{10}{\ns} - which translates into \SI{1}{\m} spatial resolution \cite{Fellay1997a}. Several methods were proposed to improve the spatial resolution of the time domain technique including dark pulses \cite{Brown2007} and Brillouin echo distributed sensing (BEDS) \cite{Thevenaz2010,Foaleng2010,Stiller2010a}, which improved the spatial resolution to a few \SI{}{\cm}. 

A more recent SBS-based sensing technique, which offers higher spatial resolution is called Brillouin optical correlation domain analysis (BOCDA). This scheme relies on the correlation between the cw pump and probe waves \cite{Hotate2000}. Different variations of BOCDA include broad-spectrum pump and probe sources based on frequency modulation \cite{Song2008a,Hasegawa1999}, random phase modulation \cite{Zadok2012,Antman2013}, filtered ASE source \cite{Cohen2014a} and chaotic laser \cite{Zhang2018e}. 
Spatial resolution of a few \SI{}{mm} in optical fiber was reported using this technique \cite{Song2006,Cohen2014a}, which unlike the time domain technique is not limited by the phonon lifetime \cite{Hotate2000}.

The higher spatial resolution offered by BOCDA opens up the opportunity to monitor and characterize smaller and more sensitive structures such as micro-fibers \cite{Chow2018} and on-chip photonic waveguides \cite{Zarifi2018}.
SBS-response characterization becomes critical in micro- and nano-scale structures, where the geometrical non-uniformities along the waveguide result in a broadening of the SBS spectrum \cite{Wolff2016} and influence the applications which rely on the narrow-linewidth of the SBS process \cite{Kabakova2013,Lee2012,Loh2015a,Otterstrom2018,Morrison2017,Marpaung2015a,Sancho2012a}. In addition, the SBS response in nano-scale waveguides is more sensitive to the complex geometrical features such as tapers, bends and on-chip gratings \cite{Merklein2015} due to the strong effect of the waveguide boundaries in the sub-wavelength regime \cite{Wolff2015,Qiu2013b}.  
Therefore, identifying the local SBS responses at these critical points gives some insight on the geometry-dependent acousto-optic interactions and provides a feedback for the design and fabrication step in order to improve the quality of the SBS waveguides.

\begin{figure}[t!]
\centering
%\fbox{
\includegraphics[width=\linewidth]{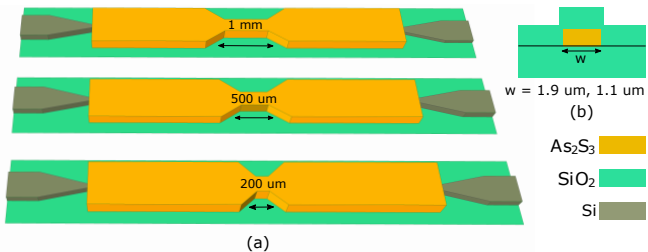}
\caption{ Schematic of the test waveguides on the silicon-chalcogenide hybrid platform. The length of the test area is reduced from \SI{1}{\mm} to \SI{200}{\um}. The inset at the top right shows the cross-section of the hybrid waveguide. The width of the waveguide ($w$) varies between \SI{1.90}{\um} to \SI{1.10}{\um} along the test waveguide.}
\label{fig:figure_1}
\end{figure}

\begin{figure}[b!]
\centering
%\fbox{
\includegraphics[width=\linewidth]{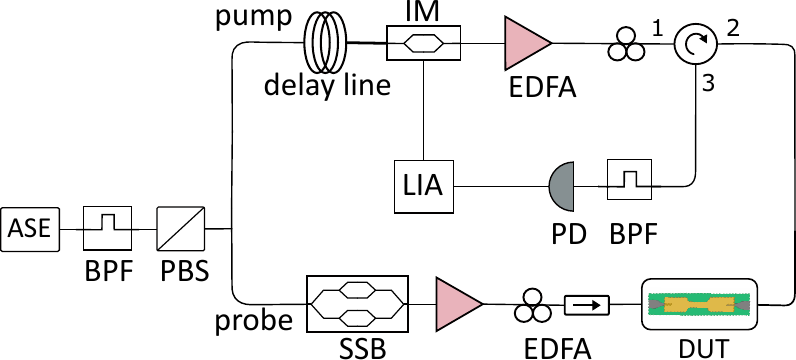}
\caption{BOCDA setup based on the ASE of an erbium doped fiber. BPF: band-pass filter, PBS: Polarisation beam splitter, SSB: single side-band modulator, IM: intensity modulator, EDFA: erbium doped fiber amplifier, DUT: device under test PD: photo detector, LIA: lock-in amplifier.}
\label{fig:figure_2}
\end{figure}

\begin{figure*}[t!]
\centering
%\fbox{
\includegraphics[width=\linewidth]{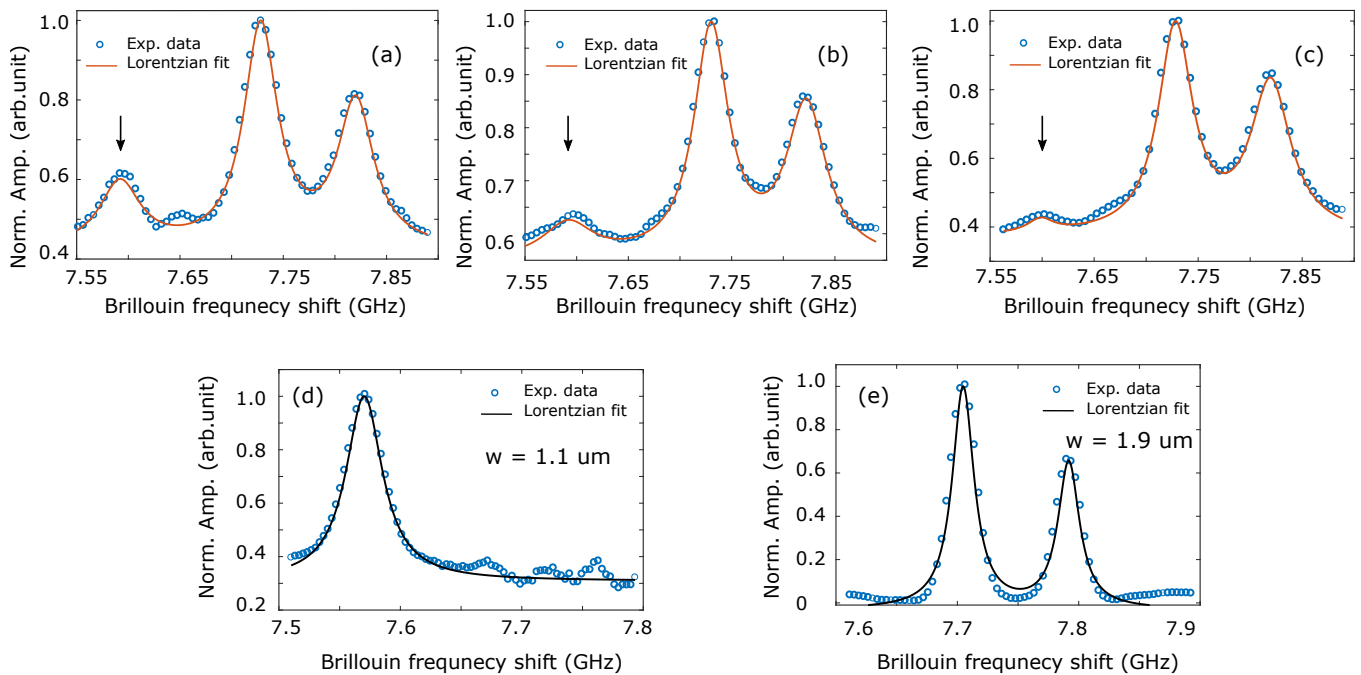}
\caption{a) Integrated SBS response of the test waveguides with (a) \SI{1}{\mm}, (b) \SI{500}{\um} and (c) \SI{200}{\um} feature size corresponding to schematic in Fig.\,\ref{fig:figure_1}. The black arrows indicates the disappearance of the peak at lower frequency attributed to the narrow waveguide region from (a) to (c). Integrated SBS response of the reference waveguides with (d) \SI{1.10}{\um} and (e) \SI{1.90}{\um} width. In all the plots, the experimental data is shown with blue dots and the Lorentzian fit is shown with solid line.}
\label{fig:figure_3}
\end{figure*}

Previous works on the local SBS response characterization in micro-scale waveguides based on BOCDA involve mapping the uniformity of a silica planar lightwave circuit (PLC) \cite{Hotate2012} and a photonic crystal fiber (PCF) \cite{Stiller2010a} and excitation and detection of surface acoustic waves in micro-fibers \cite{Chow2018}. Recently, sub-mm spatial resolution was achieved by employing the BOCDA technique in a chalcogenide photonic waveguide with an improved signal-to-noise ratio (SNR) compared to the optical fiber measurements \cite{Zarifi2018}. The chalcogenide waveguide offers a higher SBS gain due to the large opto-acoustic overlap, the higher refractive index of the core material and the smaller nonlinear opto-acoustic effective area compared to optical fibers. In addition to these advantages, the use of a phase-sensitive detection technique \cite{Hotate2000} further improves the spatial resolution and the SNR of the measurement. This experiment as the first demonstration of the sub-mm BOCDA measurement on a chip-scale, opened up the opportunity to study the effect of geometrical variation and design parameters on the overall SBS response of more complex and compact photonic-phononic waveguides with small feature such as on-chip gratings \cite{Merklein2015}. However, to monitor such small structures, the spatial resolution has to be in the order of the small waveguide features such as spiral bends, tapers and gratings, which is the focus of this work.

Following our initial reports, in this paper, we present a set of new measurements with four-fold improvement in detection capability down to \SI{200}{\um} and a comprehensive study of the geometry-dependent opto-acoustic interactions in those structures. This experiment is demonstrated in a controlled environment in which we tailor the on-chip sensing geometry to confirm the spatial resolution of the BOCDA technique. We designed SBS waveguides with width variations along their length in order to experimentally demonstrate the capability of our distributed SBS measurement system in identifying features which are rather small to be realized by an integrated SBS measurement. We further verified the experimental results with numerical calculations to explain the gain spectrum of the local opto-acoustic interactions along the photonic waveguide.
This study opens up opportunities to investigate the local SBS response of more complex structures with very fine features and is a major step forward to a better understanding of the spatial limit of opto-acoustic interactions in sub-wavelength regimes.

\section{Waveguide design and fabrication}

We designed a hybrid silicon-chalcogenide chip consisting of several waveguides, each contains a controlled width variation to characterize the local SBS responses within the waveguides and confirm the spatial resolution. 
A schematic of the test structures on a hybrid platform is shown in Fig.\,\ref{fig:figure_1} (a).
The length of the controlled width varies from \SI{1}{\mm} to \SI{200}{\um} for different waveguides. A number of reference waveguides with constant width are also designed to characterize the opto-acoustic responses at specific waveguide cross-sections. 
A cross-section view of the chalcogenide waveguide used in this work is shown in Fig.\,\ref{fig:figure_1} (b).
The hybrid silicon-chalcogenide waveguide consists of SOI grating couplers to couple the light into and out of the waveguide. The grating couplers selectively couple the fundamental TE mode into the standard single mode silicon nanowire ($\SI{450}{\nm}\times\SI{220}{\nm}$). The silicon nanowires attached to the grating couplers continue for \SI{1}{\mm} before they taper down to \SI{150}{\nm} wide tips. 
A layer of \SI{680}{\nm} thick chalcogenide ($As_{2}S_{3}$) is then deposited into the area between the two grating couplers covering the silicon tapers but leaving the grating couplers uncovered. The chalcogenide waveguides with the length of \SI{6}{\mm} are written using the electron beam lithography (EBL) technique, followed by plasma etching and are then covered by a layer of silica cladding to protect the waveguides and provide optimum acoustic confinement. 

\section{Experiment}

The local SBS response achieved in this measurement is based on BOCDA using the amplified spontaneous emission (ASE) of an erbium doped fiber. This technique was first employed to measure a local hot spot in an optical fiber with \SI{4}{\mm} spatial resolution \cite{Cohen2014a}. In this technique, the polarized ASE source provides a highly uncorrelated source for the pump and probe waves. The degree of the correlation between the pump and the probe is controlled by the ASE bandwidth. Increasing the ASE bandwidth reduces the correlation between the pump and the probe waves and results in a narrow correlation peak in time. The duration of the correlation peak determines the spatial resolution of the measurement, which can be approximated by: $\frac{1}{2}V_{g}\Delta{t}$ with $V_{g}$ being the group velocity and $\Delta{t}$ is inversely related to the ASE bandwidth \cite{Cohen2014a}. 
As the correlation peak becomes shorter in time, the signal from the local SBS interaction becomes weaker compared to the background noise from the spontaneous scattering at all the other points in the medium (outside the correlation peak). Therefore, the SNR decreases and sets a lower limit on the practical spatial resolution of the BOCDA measurement technique. In addition, for ASE bandwidths larger than the Brillouin frequency shift (BFS), the separation between the back-reflected pump and the amplified probe becomes challenging due to the large spectral overlap \cite{Zarifi2018}. In our setup, we use $As_{2}S_{3}$ strip waveguides with high SBS gain coefficient and a lock-in amplifier (LIA) to improve the SNR and obtain spatial resolutions beyond the limits achieved in the optical fiber measurement.

A schematic of the experimental setup is presented in Fig.\,\ref{fig:figure_2}. It consists of the ASE source whose bandwidth is controlled by a tunable band-pass filter (BPF). It then divides into the counter-propagating pump and probe arms. The probe wave goes through a single side-band modulation using a dual-parallel Mach-Zehnder modulator (DPMZM) with a carrier suppression of 20 dB and a side-band suppression of 15 dB. The pump wave is intensity-modulated with pulse lengths of \SI{500}{\ns} and a frequency of \SI{100}{\kHz} and is synchronized with the LIA. The light is coupled in and out of the waveguide using silicon grating couplers with the measured total back-reflection of \SI{-18}{\dB}. A sharp optical filter is added before the photo detector to remove the pump back-reflection as much as possible before sending the measured signal to the LIA. The essential part of this experiment is cutting the pump back-reflection, because by filtering out the pump back-reflection, part of the SBS signal will also be removed. However, we achieved enough signal even after cutting \SI{90}{\percent} of the response to detect a segment of \SI{200}{\um} of the waveguide.

\begin{figure*}[t]
\centering
%\fbox{
\includegraphics[width=\linewidth]{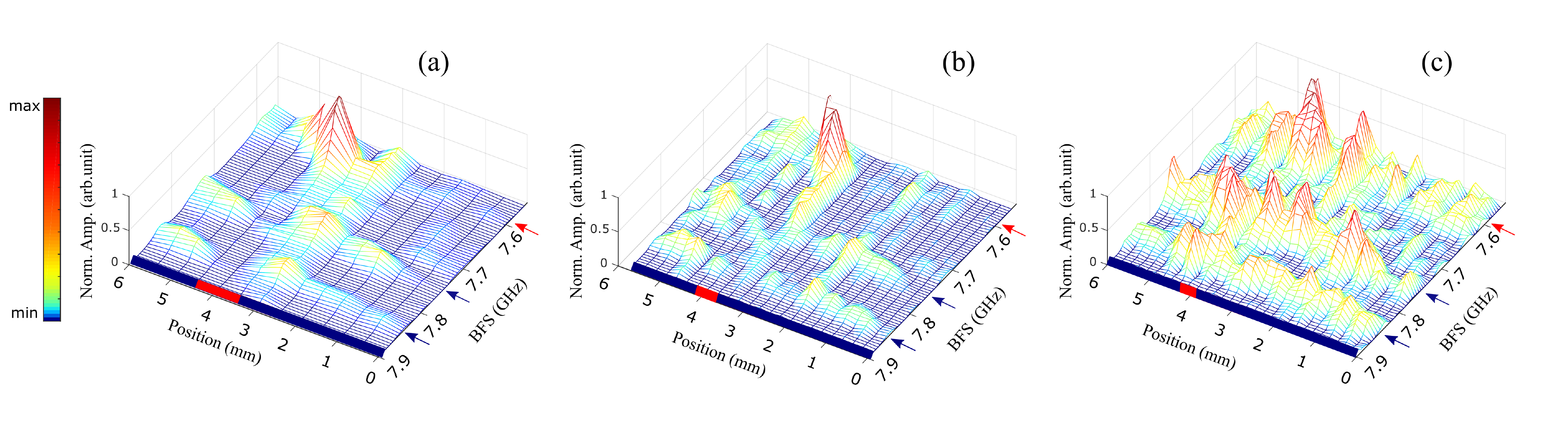}
\caption{BOCDA response of the test waveguide with a) \SI{1}{\mm} feature size and \SI{1}{\mm} spatial resolution, b) \SI{500}{\um} feature size and \SI{750}{\um} spatial resolution and c) \SI{200}{\um} feature size and \SI{750}{\um} spatial resolution. The approximate location of the narrow-width region is shown in red color along the position axis. The blue and red arrows along the BFS axis indicate the BFS in the wide and narrow waveguide regions, respectively. The vertical axis in all plots shows the normalized amplitude (Norm.Amp.).}
\label{fig:figure_4}
\end{figure*}

In order to identify the SBS response of the entire test waveguides, we first performed an integrated SBS measurement by setting the filter bandwidth to \SI{3}{\GHz}, corresponding to \SI{12}{\mm} spatial resolution (twice the length of the waveguide). \mbox{Fig.\,\ref{fig:figure_3} (a)-(c)} show the integrated SBS measurement of the test waveguides corresponding to the structures shown in Fig.\,\ref{fig:figure_1} (a). 
The waveguide with \SI{1}{\mm} feature size has three peaks in the Brillouin gain spectrum namely at \SI{7.59}{\GHz}, \SI{7.72}{\GHz} and \SI{7.81}{\GHz}. 
As the feature size reduces from \SI{1}{\mm} to \SI{200}{\um}, the peak at lower frequency (\SI{7.59}{\GHz}) disappears. 
These measurements show that there is a link between the lower frequency peak and the narrow-width feature in the middle of the test waveguides. To further confirm this measurement, we investigate the integrated SBS response of two reference waveguides with \SI{1.10}{\um} and \SI{1.90}{\um} constant width using a cw laser as the pump and the probe source. The result of these measurements are plotted in Fig.\,\ref{fig:figure_3} (d) and (e). As it is shown in these plots, the \SI{1.10}{\um}-wide waveguide has a BFS at \SI{7.57}{\GHz} and the \SI{1.90}{\um}-wide waveguide has a double peak profile at \SI{7.70}{\GHz} and \SI{7.79}{\GHz}, which is in agreement  with the measurements shown in Fig.\,\ref{fig:figure_3} (a) to (c). The slight offset between the BFS observed in the two sets of measurements (with \SI{3}{\GHz} ASE bandwidth and the cw laser) is due to the fact that the center frequency of the two sources were slightly different. 
% the laser measurement is shifted by 20 MHz toward lower frequencies.

In order to measure the longitudinal feature in the first test waveguide, a distributed SBS measurement was performed by setting the ASE bandwidth to \SI{60}{\GHz}, corresponding to \SI{1}{\mm} spatial resolution in the waveguide. The delay line is set such that the optical length of the pump and the probe arms are equal and then by stepping the delay line with \SI{500}{\um} steps, the waveguide with \SI{1}{\mm} feature size was scanned. 
The result of this measurement is shown in Fig.\,\ref{fig:figure_4} (a), where the BFS peak at \SI{7.59}{\GHz} appears at position \SI{3.5}{\mm} and disappears at \SI{4.5}{\mm}. Outside this region, the Brillouin gain spectrum mainly shows two peaks at \SI{7.72}{\GHz} and \SI{7.81}{\GHz} which indicates that the correlation peak is scanning the wider section of the waveguide. 
The ASE bandwidth is then increased to \SI{80}{\GHz}, corresponding to a spatial resolution of \SI{750}{\um} to detect the \SI{500}{\um} feature in the second test waveguide with step sizes of \SI{250}{\um}, as presented in Fig.\,\ref{fig:figure_4} (b). As it is seen in this plot, the BFS peak at \SI{7.59}{\GHz} appears at position \SI{3.75}{\mm} and disappears at position \SI{4.25}{\mm}. 
Lastly, the waveguide with \SI{200}{\um} feature was measured using the same ASE bandwidth (\SI{80}{\GHz}), however the step size is now reduced to \SI{200}{\um}. This measurement is shown in Fig.\,\ref{fig:figure_4} (c), where the \SI{7.59}{\GHz} peak appears between the positions \SI{3.8}{\mm} and \SI{4.2}{\mm} and has the highest amplitude at position \SI{4.0}{\mm}. 
As it is seen in this figure, the quality of the detected local signals deteriorates in the last measurement due to the lower spatial resolution compared to the feature size and also the lower SNR. Some residual of the \SI{7.59}{\GHz} peak could be observed in traces away from the feature, which is due to the weak SNR and the fact that the signal level is now close to the background noise from the spontaneous Brillouin scattering which happens outside the correlation peak.  

In addition, in Fig.\,\ref{fig:figure_4} (a) and (b), the intensity of the local SBS response at the narrow region waveguide is higher compared to the wide region. This is due to the fact that the SBS gain coefficient per unit length defined as $\frac{g_{B}}{A_{\textup{eff}}}$, with $g_{B}$ being the SBS gain coefficient, is inversely related to the waveguide's effective opto-acoustic interaction area $A_{\textup{eff}}$. Therefore, the SBS gain in the narrower waveguide with approximately four times smaller cross-section, is expected to be stronger than in the wider waveguide. This is confirmed by the normalized Brillouin gain spectrum obtained from the experiment and shown in Fig.\,\ref{fig:figure_4} (a) to (c), where the peak amplitudes at frequencies \SI{7.72}{\GHz} and \SI{7.81}{\GHz} are approximately \SI{25}{\percent} less than the peak amplitude at frequency \SI{7.59}{\GHz}.

\section{Simulation and analysis}

The dependency of the BFS ($\nu_{B}$) on material properties and effective refractive index can be approximated by Eq.\,\ref{eq:3}.

\begin{figure*}[t!]
\centering
%\fbox{
\includegraphics[width=\linewidth]{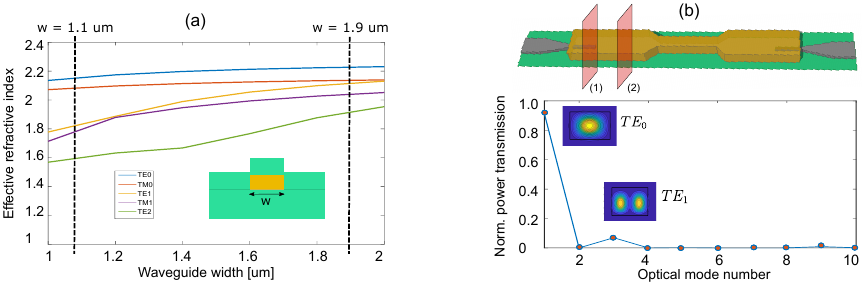}
\caption{(a) Schematic of the chalcogenide waveguide cross-section and its dispersion plot, showing the guided optical modes at the two  waveguide widths of \SI{1.10}{\um} and \SI{1.90}{\um}. (b) Optical mode transition from the silicon taper (plate 1) to the chalcogenide waveguide (plate 2). Bottom: Transmitted optical power from the silicon nanowire to the chalcogenide waveguide through the silicon taper. \SI{90}{\percent} of the optical power is transmitted to the fundamental TE mode and less than \SI{10}{\percent} of the power is in the $TE_{1}$ mode.}
\label{fig:figure_5}
\end{figure*}

\begin{equation}
\nu_{B} = \frac{2n_{\textup{eff}}V_{a}}{\lambda_{p}},
\label{eq:3}
\end{equation}
where $n_{\textup{eff}}$ is the effective refractive index, $V_a$ is the acoustic mode velocity and $\lambda_{p}$ is the pump wavelength.
This relation is valid under the assumption that the waveguide dimensions are much larger than the acoustic wavelength ($w , h \gg \frac{2\pi V_{a}}{\Omega}$) \cite{Poulton2013a}, where $w$ and $h$ stand for the waveguide  width and thickness, respectively and $\Omega$ is the acoustic angular frequency. This assumption is correct for the optical fibers, however for the sub-wavelength and wavelength-scale waveguides the medium can no longer considered isotropic and the optical field will have components in the direction of propagation which will affect the effective refractive index term \cite{Poulton2013a}. Furthermore, it was demonstrated that in nanostructures other forces 
such as radiation pressure influence the acoustic mode excitations \cite{Rakich2012}. In order to include all these effects in our study, we performed a fully vectorial analysis to calculate the BFS in the chalcogenide structures following the approach presented in \cite{Wolff2015}. However, for our hybrid chalcogenide structures the effect of radiation pressure on the backward SBS gain turns out to be negligible, which is confirmed through numerical calculations.
Different waveguide cross-sections accommodate different optical and acoustic modes and therefore the effective opto-acoustic overlap is different from one waveguide geometry to the other. The change in the waveguide cross-section, therefore, manifests in the Brillouin gain spectrum and shifts the BFS, which is what we measure in this experiment. 

The chalcogenide waveguide with cross-sections shown in the inset of Fig.\,\ref{fig:figure_1} (b) supports multiple optical modes as well as acoustic modes. Therefore, careful design considerations are required in order to precisely excite the correct optical mode and comprehensive analysis needs to be done in order to identify the acoustic modes involved in the SBS interaction. The dispersion plot of the chalcogenide waveguide is shown in Fig.\,\ref{fig:figure_5} (a). The single mode operation of the chalcogenide waveguide is achieved via the adiabatic silicon taper.
The silicon tapers are long enough (\SI{100}{\um}) to guarantee a smooth optical mode transition from the silicon nanowire to the chalcogenide waveguide \cite{Casas-Bedoya2016}. The fundamental TE mode transition through the adiabatic taper is simulated using a commercial-grade simulator based on the finite-difference time-domain method \cite{Lumerical:2009:Misc}, presented in Fig.\,\ref{fig:figure_5} (b). As presented in this figure, the majority of the optical power is coupled into the fundamental TE mode of the chalcogenide waveguide and less than \SI{10}{\percent} of the power is transmitted into the higher order TE mode. Therefore, although the waveguide could support multiple optical modes, the silicon tapers were designed to selectively excites only the fundamental TE mode.

\begin{figure}[b]
\centering
%\fbox{
\includegraphics[width=\linewidth]{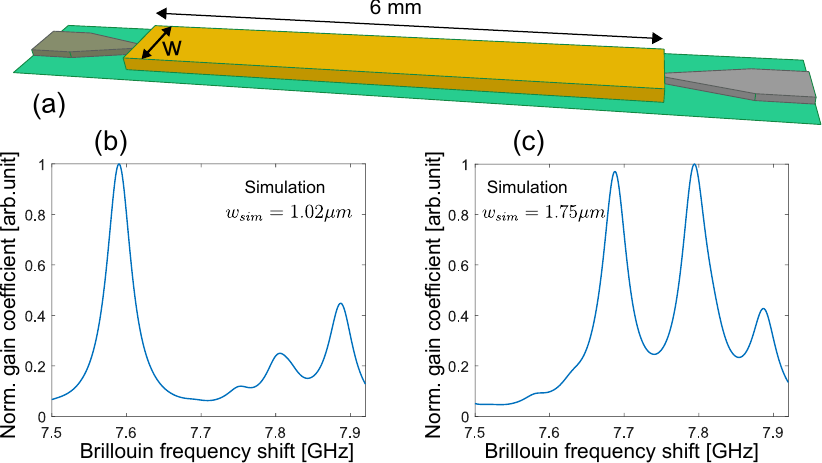}
\caption{a) Schematic of a reference waveguide with constant width. Calculated normalized Brillouin gain spectrum for $h_{sim} = \SI{707}{\nm}$ thick waveguide with b) $w_{sim} = \SI{1.02}{\um}$ and c) $w_{sim} = \SI{1.75}{\um}$ width.
}
\label{fig:figure_6}
\end{figure}

The opto-acoustic response is calculated using COMSOL Multiphysics software \cite{COMSOL:2009:Misc} after ref.\cite{Wolff2015}. We calculated the overlap between the fundamental optical TE mode with the acoustic modes present in the frequency span between \SI{7.50}{\GHz} and \SI{7.90}{\GHz} and reconstructed the Lorentzian Brillouin gain spectrum based on the strength of the opto-acoustic overlap within this spectrum.
The simulated normalized Brillouin gain spectrum for the two reference waveguide geometries are shown in Fig.\, \ref{fig:figure_6}. Fabrication variations in thickness and width of \SI{7}{\percent} were allowed in this calculation using a corner analysis \cite{Chrostowski2013} to find a reasonable match between the simulation and the experiment. 
The simulation plots shown in Fig.\,\ref{fig:figure_6} (b) and (c) represent waveguides with the simulation thickness ($h_{sim}$) of \SI{707}{\nm} and the simulation widths ($w_{sim}$) of \SI{1.02}{\um} and \SI{1.75}{\um}, respectively.
The material properties including stiffness tensor coefficients, density and photoelastic tensor coefficients for this simulation were set after Ref. \cite{Smith2016}. Comparing Fig.\,\ref{fig:figure_6} (b)-(c) with the experimental result shown in Fig.\,\ref{fig:figure_3} (d)-(e), we find a good agreement between the experiment and the simulation. This confirms that the shift in the peak of the Brillouin spectrum profile is an effect of the waveguide geometry and the double peak profile observed in the wider waveguide is a result of the existence of two or more higher order acoustic modes which have strong overlap with the optical fundamental TE mode in the vicinity of the BFS. 

%\textcolor{red}{The acoustic mode profiles contributing to the peaks of the calculated Brillouin gain spectrum at the two waveguide geometries with \SI{1.02}{\um} and \SI{1.75}{\um} are shown in Fig.\,\ref{fig:figure_2} (d) and (e), respectively.}

\section{Discussion}

By comparing Fig.\,\ref{fig:figure_3} (d) and (e) with Fig.\,\ref{fig:figure_6} (b) and (c), a good agreement between the  measured integrated SBS response and the simulation is observed. However, an additional BFS peak at \SI{7.90}{\GHz} is observed in the simulation but not captured in the experiment. This is most likely due to the fact that the SNR of the measurement was not high enough to detect the rather weak SBS response which is generated by some higher order acoustic modes.

When simulating the optical mode transition from the silicon nanowire to the chalcogenide waveguide, we considered up to \SI{50}{\nm} vertical offset of the taper tip from the center of the chalcogenide waveguide as well as \SI{20}{\percent} variations of the taper width in order to include the effect of fabrication variations.
The effect of the taper misalignment and taper tip width variation on the optical mode transition were negligible and the optical mode transition profiles overlap very closely with the plot shown in Fig.\,\ref{fig:figure_5} (b), therefore they are not plotted here.
In addition, as it is plotted in Fig.\,\ref{fig:figure_5} (b), approximately \SI{10}{\percent} of the transmitted power from the silicon nanowire is coupled into the $\textup{TE}_{1}$ mode of the \SI{1.90}{\um}-wide chalcogenide waveguide, which could be considered as the origin of the second peak appeared at the lower frequency in the Brillouin gain spectrum. We examined this possibility by calculating the opto-acoustic overlap between the $\textup{TE}_{1}$ mode and the acoustic modes in the waveguide and found out that the frequency splitting between the two peaks in this case is at least \SI{200}{\MHz}, which does not match the experiment. We further studied the contribution of the $\textup{TM}_{0}$ mode in the opto-acoustic overlap to investigate the possibility of mode coupling within the waveguide. The opto-acoustic overlap between the $\textup{TM}_{0}$ mode and the acoustic modes of the waveguide resulted in a frequency splitting of \SI{180}{\MHz}, which is larger than what we observed in the experiment.
%To further verify the origin of the double peak in the wide waveguide, the grating couplers were removed by cleaving the chip so that both TE and TM modes were allowed to couple into the waveguide via a polarization controller. By chanting the polarization from one state to the other, the double peak profile persists, however the intensity of the peaks varied at different polarizations. This last experiment indicates that the polarization has an impact on the opto-acoustic overlap, however since none of the peaks vanishes entirely by changing from TE to TM mode, we can confirm that the double peak profile is not due to the existence of multiple optical modes but is mainly caused by the acoustic modes.  

Finally, as it was mentioned earlier, the limiting factor in this experiment was the wide overlap between the back-reflected pump and the amplified probe. Improving fabrication techniques such as the use of tilted grating couplers \cite{Li2013a} with very low back reflection can improve the SNR and allows for higher spatial resolution measurement.

\section{Conclusion}

In this work, we reported four-fold improvement in detection capability of BOCDA measurement compared to previous works \cite{Zarifi2018a}. In addition, we performed numerical simulation to explain the interaction between the optical and acoustic modes at different waveguide cross-sections.
This setup provides a powerful platform to test and measure local opto-acoustic responses within the micro- and nano-wires with sub-mm feature size. Moreover, by further increasing the spatial resolution, this technique could provide valuable information regarding the spatial limits of the nonlinear opto-acoustic interaction within the medium.

% Note that \emph{Optics Letters} uses an abbreviated reference style. Citations to journal articles should omit the article title and final page number; this abbreviated reference style is produced automatically when the \emph{Optics Letters} journal option is selected in the template, if you are using a .bib file for your references.

% However, full references (to aid the editor and reviewers) must be included as well on a fifth informational page that will not count against page length; again this will be produced automatically if you are using a .bib file.

% \bigskip
% \noindent Add citations manually or use BibTeX. See \cite{Zhang:14,OSA,FORSTER2007,testthesis}.

% Bibliography

 \section*{Funding Information}

This work was funded by the Australian Research Council (ARC) Laureate Fellowship (FL120100029) and the Centre of Excellence program (CUDOS CE110001018).

This work was performed in part at the Melbourne Centre for Nanofabrication (MCN) and the RMIT Micro Nano Research Facility (MNRF) in the Victorian Node of the Australian National Fabrication Facility (ANFF).

We acknowledge the joint grant from the Max Planck Society and the Fraunhofer Society (PowerQuant).

%\section{References}

%\bigskip
%\noindent Add citations manually or use BibTeX. See \cite{Zhang:14,OSA,FORSTER2007,testthesis}.

% Bibliography
%\bibliography{OFS_Ref}

\bibliographystyle{ieeetr}
\bibliography{JOSAB_REF}

%\begin{thebibliography}{99}

%\end{thebibliography}

\end{document}